\documentclass[preprint,chaos,aps,amssymb,amsmath,floatfix]{revtex4}
\usepackage{graphicx}

\newcommand{\be}{\begin{equation}}
\newcommand{\ee}{\end{equation}}
\newcommand{\bi}[1]{\vspace{-3mm} \bibitem{#1}}

\begin{document}

\title{Dynamics of the Chain of Oscillators
with Long-Range Interaction: From Synchronization to Chaos}

\author{G.M. Zaslavsky}
\affiliation{Courant Institute of Mathematical Sciences,
New York University,
251 Mercer St., New York, NY 10012, USA\\
Department of Physics, New York University,
2-4 Washington Place, New York, NY 10003, USA}
\author{M. Edelman}
\affiliation{Courant Institute of Mathematical Sciences,
New York University,
251 Mercer St., New York, NY 10012, USA}
\author{V.E. Tarasov}
\affiliation{Courant Institute of Mathematical Sciences,
New York University,
251 Mercer St., New York, NY 10012, USA\\
Skobeltsyn Institute of Nuclear Physics, 
Moscow State University, Moscow 119992, Russia}

\date{\today}

\begin{abstract}
We consider a chain of nonlinear oscillators with long-range interaction
of the type $1/l^{1+\alpha }$, where $l$ is a distance between oscillators
and $0< \alpha <2$. In the continues limit the system's dynamics is
described by the Ginzburg-Landau equation with complex coefficients. Such
a system has a new parameter $\alpha$ that is responsible for the
complexity of the medium and that strongly influences possible regimes of
the dynamics. We study different spatial-temporal patterns of the dynamics
depending on $\alpha$ and show transitions from synchronization of the
motion to broad-spectrum oscillations and to chaos.
\end{abstract}

\maketitle

{\bf A chain of nonlinear interacting oscillators is a model of 
the wide-spread investigation of different physical phenomena such as
synchronized behavior of the system, bifurcations to different regimes,
spatial-temporal turbulent or chaotic dynamics, different instabilities,
appearance of defects, and many others. The long range interaction between
oscillators leads to a new qualitative dynamics and thermodynamics. Our
consideration is related to the power-law long-range interaction that
makes it possible to find new regimes of the system and to establish a link
of the corresponding equations of motion to the equations with fractional
derivatives.}

\section{Introduction}

The goal of the paper is to consider different dynamical regimes,
from synchronization to chaos and turbulence, 
in a chain of large number of coupled nonlinear oscillators
with long-range interaction (LRI) of a power type.
The potential of interaction is nonlocal and proportional
to $1/l^{1+\alpha}$ with $l$ as a distance between oscillators and
$0<\alpha<2$, ($\alpha \ne 1$).
It will be called $\alpha$-interaction.
Transitions between different regimes of the chain behavior 
are considered as a function of $\alpha$.
In the continuous limit the system reduces to the fractional
generalization of the Ginzburg-Landau (FGL) equation 
and the chain of oscillators can be considered as a discretized
model of FGL, called DFGL.
Since we also consider the model with complex coefficients,
it either can be related to a fractional generalization of
the complex nonlinear Schr\"odinger equation. 

The literature related to this type of problem is fairly wast 
and we would refer only to some reviews and closely related articles.
The complex Ginzburg-Landau (CGL) equation
is typically considered for pattern formation in different media
\cite{Z1,AK,Z2}.
CGL equation appears in numerous physical models 
(see for example \cite{TK,Mikh,Kur3,Kur4}).
Long-range interaction with finite radius of interaction
was considered for complex media in \cite{Kur1,Z3}
and for the $\alpha$-interaction in \cite{Dyson,Z4,Ruffo,TCT,Z5}.
Different regimes of synchronization in the chain of 
coupled oscillators with nearest neighbour interaction
can be found in reviews \cite{Pik1,BKOVZ} while the synchronization
with $\alpha$-interaction for DFGL was found in \cite{TZ3,KZ}.
The FGL equation was introduced in \cite{WZ} to describe 
the wave propagation in complex media, when 
the dispersion law has fractional power of the wave number.
The related case appears in description of weak turbulence \cite{Z6}.
More examples related to the chains with LRI and their solution and
breather type solutions and applications can be found in \cite{Z7,Z8}.

The main part of this paper is 
a numerical simulation of the chain of $N=512$ coupled oscillators
with $\alpha$-interaction and constant external force
that pump energy into the system.
This system links to the FGL equation in continuous limit 
based on the formal procedure introduced in \cite{LZ,TZ3}.

It will be shown that the forcing oscillatory media
with LRI exhibits transition to chaos and turbulence 
when $\alpha$ decreases.
As another interesting regime, 
we found a "collectivized" limit cycle that 
has a broad spectrum in phase space of any individual oscillator.
Such a cycle exists within an interval of $\alpha$.

In Sec. 2, we present the basic equation for coupled oscillators
and their continuous media limit.
This part includes some modified version of 
the results of \cite{TZ3,Z9}.
In Sec. 3, we consider fractional generalization of 
the CGL equation and some of its stability criteria.
The Secs4-6 play an auxiliary role to identify
an interesting domain of parameters to be studied.
The main numerical results are presented in Secs. 7-9.
It is important to note that the main studied object
is the chain of nonlinear oscillators, and 
a part of the obtained results can be appropriate 
for the complex FGL equation at least for a finite time. 
The FGL equation can also be used for 
some estimates of the critical parameters to
change the regimes of behavior of the chain of oscillators.

\section{Chain of oscillators with long-range interaction}

Following \cite{Z9}, let us introduce the system by its action
\be \label{Sun}
 S[Z_n]=\int^{+\infty}_{-\infty} dt  \Bigl(
\sum_{n=-\infty}^{+\infty}
\left[ \frac{1}{2} \;  \dot{Z}_n(t) \dot{Z}_n(t)
-V (Z_n(t)) \right] 
-\sum_{\substack{n, m=-\infty \\ m \ne n}}^{+\infty} \; 
U (Z_n(t),Z_m(t)) \Bigr) ,
\ee
where $Z_n$ are displacements of the oscillators from their equilibrium, 
and $U (Z_n(t),Z_m(t))$ is defined by
\be \label{UU}
U(Z_n(t),Z_m(t))= \frac{1}{4} g_0 J_{\alpha}(|n-m|) \, (Z_n(t)-Z_m(t))^2
, \   \ (n \ne m)
\ee
with $g_0$ as an interaction constant, and with
interparticle interaction 
\be \label{189}
J_{\alpha}(|n-m|) =\frac{1}{|n-m|^{\alpha+1}}, \quad (\alpha>0) .
\ee
The usual linear oscillator term $Z_n^2(t)$ can be included into
$V(Z_n(t))$. The corresponding Euler-Lagrange equations are
\be \label{Z2a}
\frac{dZ_n }{dt}+
g_0 \sum_{\substack{m=-\infty \\ m \ne n}}^{+\infty} \; 
J_{\alpha}(|n-m|) \; [Z_m(t)-Z_n(t)] + \; F (Z_n(t))=0 ,
\ee
where $F(u)=\partial V (u)/ \partial u$.


Continuous limit of equation (\ref{Z2a}) can be defined by
a transform operation from $Z_n(t)$ to $Z(x,t)$
\cite{LZ,TZ3}. 
Firstly, define $Z_n(t)$ as Fourier coefficients of some 
function $\hat{Z}(k,t)$, $k \in [-K_0/2, K_0/2]$, i.e.
\be \label{ukt}
\hat{Z}(t,k) = \sum_{n=-\infty}^{+\infty} \; Z_n(t) \; e^{-i k x_n} =
{\cal F}_{\Delta} \{Z_n(t)\} ,
\ee
where  $x_n = n \Delta x$, and $\Delta x=2\pi/K_0$ is a
distance between nearest particles in the chain, and
\be \label{un} 
Z_n(t) = \frac{1}{K_0} \int_{-K_0/2}^{+K_0/2} dk \ \hat{Z}(t,k) \; e^{i k x_n}= 
{\cal F}^{-1}_{\Delta} \{ \hat{Z}(t,k) \} . 
\ee
Secondly, in the limit $\Delta x \rightarrow 0$ ($K_0 \rightarrow \infty$) 
replace $Z_n(t)=(2\pi/K_0) Z(x_n,t) \rightarrow  Z(x,t) dx$, 
and  $x_n=n\Delta x= 2\pi n/K_0 \rightarrow x$.
In this limit, Eqs. (\ref{ukt}), (\ref{un}) are transformed into 
the integrals
\be \label{ukt2} 
\tilde{Z}(t,k)=\int^{+\infty}_{-\infty} dx \ e^{-ikx} Z(t,x) = 
{\cal F} \{ Z(t,x) \} = \lim_{\Delta x \rightarrow 0} {\cal F}_{\Delta} \{Z_n(t)\} , 
\ee
\be \label{uxt}
Z(t,x)=\frac{1}{2\pi} \int^{+\infty}_{-\infty} dk \ e^{ikx} \tilde{Z}(t,k) =
 {\cal F}^{-1} \{ \tilde{Z}(t,k) \}= 
\lim_{\Delta x \rightarrow 0} {\cal F}^{-1}_{\Delta} \{ \hat{Z}(t,k) \}  . 
\ee

Applying (\ref{ukt}) to (\ref{Z2a})
and performing the limit (\ref{ukt2}), we obtain
\be \label{200}
\frac{\partial Z(t,x)}{\partial t}+
g_{\alpha} \frac{\partial^{\alpha} Z(t,x)}{\partial |x|^{\alpha}}+
F(Z(t,x))=0, \quad (1<\alpha<2) ,
\ee
where $ \partial^{\alpha} / \partial |x|^{\alpha}$ is 
the fractional Riesz derivative \cite{KST} defined by
\[ \frac{\partial^{\alpha} Z(t,x)}{\partial |x|^{\alpha}}=
{\cal F}^{-1}\{ |k|^{\alpha} \tilde Z(t,k) \} ,  \]
and
\be
g_{\alpha}=2 g_0 (\Delta x)^{\alpha}  \Gamma(-\alpha) \cos 
\left( \frac{\pi \alpha}{2} \right)
\ee
is the renormalized interaction constant. 
Equations of type (\ref{200}) with different nonlinear terms
were considered in \cite{LZ,TZ3}. 
For other values of $\alpha$ one can obtain 
similar equation to (\ref{200})
by performing the corresponding transform operation.


\section{Complex nonstationary Ginzburg-Landau equation }

Consider a chain of the nearest-neighbour
coupled nonlinear oscillators described by the equations 
\be
\frac{d}{dt} Z_n(t)=(1+ia)Z_n -(1+ib)|Z_n|^2z_n+
(c_1+ic_2)(Z_{n+1}-2Z_{n}+Z_{n-1}),
\ee
where we assume that all oscillators have the same internal parameters.
Such equations can be obtained from  (\ref{Z2a}) for $\alpha=\infty$ and
the corresponding choice of $g_0$ and $V(Z_n)$.
Transition to the continuous medium assumes 
that the difference $Z_{n+1}-Z_n$ is of the same order as $\Delta x$, 
and the interaction constants $c_1$ and $c_2$ are fairly large.
Setting $c_1=g (\Delta x)^{-2}$, $c_2=gc(\Delta x)^{-2}$
with new constants $g$ and $c$, we get
\be \label{A1}
\frac{\partial}{\partial t}Z=(1+ia)Z-(1+ib)|Z|^2Z+
g(1+ic)\frac{\partial^2}{\partial x^2} Z ,
\ee
which is a time-dependent complex Ginzburg-Landau (CGL) equation.

Let us come back to the equation for nonlinear oscillators 
with fractional long-range coupling 
\be \label{D11}
\frac{d}{dt}Z_n=(1+ia)Z_n-(1+ib)|Z_n|^2 Z_n +
g_0 \sum_{m \not=n} \frac{1}{|n-m|^{\alpha+1}} (Z_n-Z_m) ,
\ee
where $Z_n=Z_n(t)$ is the position of the $n$-th oscillator 
in the complex plane, $1<\alpha<2$, and the corresponding choice of 
$F(Z_n)=(1+i\alpha )Z_n- (1+ib )|Z_n^2|Z_n$ has neen done.
The corresponding equation in the continuous limit is
\be \label{A2}
\frac{\partial}{\partial t}Z=(1+ia)Z-(1+ib)|Z|^2Z+
g(1+ic) \frac{\partial^{\alpha}}{\partial |x|^{\alpha}} Z ,
\quad (1<\alpha<2, \  \ \alpha \ne 1)
\ee
where $g(1+ic)=g_0 a_{\alpha}$,
\be \label{D5}
a_{\alpha} =2\Gamma(-\alpha)\cos(\pi \alpha/2) .
\ee
Eq. (\ref{A2}) is a fractional generalization of the complex 
nonstationary Ginzburg-Landau (FGL) equation (\ref{A1}).
Equation (\ref{D11}) is the same as (\ref{Z2a}) with a special choice of $V$.
It was suggested in \cite{WZ} 
to describe complex media with fractional dispersion law.

The FGL equation (\ref{A2}) can be presented as the system of real equations 
\[
\frac{d}{dt} X=\Bigl(1-g  \frac{\partial^{\alpha}}{\partial |x|^{\alpha}} \Bigr)X -
\Bigl(a-gc \frac{\partial^{\alpha}}{\partial |x|^{\alpha}} \Bigr)Y-(X^2+Y^2)(X-bY) , 
\]
\be \label{B9c}
\frac{d}{dt} Y=\Bigl(1-g  \frac{\partial^{\alpha}}{\partial |x|^{\alpha}} \Bigr)Y +
\Bigl(a-gc \frac{\partial^{\alpha}}{\partial |x|^{\alpha}} \Bigr)X-(X^2+Y^2)(Y+bX) ,
\ee 
where $X=X(t,x)$  and $Y=Y(t,x)$ are real and imaginary parts of $Z(t,x)$.

The results of numerical simulation of the system (\ref{D11})
can be applied for a finite time for the FGL equation (\ref{A2})
although it is well known that typically the discrete systems are more chaotic
than their continuous counterpart.
The system (\ref{D11})
will be called discretized FGL equation or simply DFGL.

\section{Parameters of stability of plane-wave solution}

In this section, we obtain some conditions of instability
that will be implemented in numerical simulations 
of the system (\ref{D11}). These conditions are easy to derive and
interpret considering the FGL equation.

To obtain a particular solution of the FGL equation (\ref{A2}) 
with a fixed wave number $K$, we consider $Z$ in the form 
\be \label{A6}
Z(x,t)=A(K,t)e^{iKx} .
\ee
Substitution of (\ref{A6}) into (\ref{A2}) gives
\be \label{A8}
\frac{\partial}{\partial t}A(K,t)=(1+ia)A-(1+ib)|A|^2A-
g(1+ic)|K|^{\alpha}A .
\ee
The plane-wave solution of (\ref{A8}) is
\be \label{B6}
A(x,t)=(1-g|K|^{\alpha})^{1/2} e^{iKx-i\omega_{\alpha}(K)t} , 
\quad 1-g|K|^{\alpha}>0 ,
\ee
where
\be \label{B7}
\omega_{\alpha}(K)=(b-a)+(c-b)g|K|^{\alpha} ,
\quad 1-g|K|^{\alpha}>0 . 
\ee
Solution of (\ref{B6}) can be presented as
\[ X_0(x,t)=(1-g|K|^{\alpha})^{1/2} 
\cos \left( Kx-\omega_{\alpha}(K)t+\theta_0 \right) , \]
\be \label{PW}
Y_0(x,t)=(1-g|K|^{\alpha})^{1/2} \sin \left( Kx-\omega_{\alpha}(K)t+\theta_0 \right) , 
\quad 1-g|K|^{\alpha}>0 ,
\ee
where $X=X(x,t)=\mathrm{Re} Z(x,t)$,  $Y=Y(x,t)=\mathrm{Im} Z(x,t)$, and
$\theta_0$ is an arbitrary constant phase. 
These solution can be interpreted as synchronized state of the oscillatory 
medium. In fact, it will be shown by simulation that the synchronized
solution exists also for the DFGL equation (\ref{D11}). 

To obtain the stability condition, 
consider the variation of (\ref{A8}) near solution (\ref{PW}):
\be \label{S1}
\frac{d}{dt} \delta X=A_{11}\delta X+A_{12}\delta Y, \quad
\frac{d}{dt} \delta Y=A_{21}\delta X+A_{22} \delta Y ,
\ee
where $\delta X$ and $\delta Y$ are small variations of $X$ and $Y$, and
\[ A_{11}=1-g|K|^{\alpha} -2X_0(X_0-bY_0)-(X^2_0+Y^2_0), \]
\[ A_{12}=-a+gc|K|^{\alpha} -2Y_0(X_0-bY_0)+b(X^2_0+Y^2_0), \]
\[ A_{21}=a-gc|K|^{\alpha} -2X_0(Y_0+bX_0)-b(X^2_0+Y^2_0), \]
\be \label{S2}
A_{22}=1-g|K|^{\alpha} -2Y_0(Y_0+bX_0)-(X^2_0+Y^2_0) . 
\ee
The conditions of asymptotic stability for (\ref{S1}) are
\be \label{Hur}
A_{11}+A_{22}<0 , \quad A_{11}A_{22}-A_{12}A_{21}<0 .
\ee 
Substitution of Eqs. (\ref{PW}) and (\ref{S2}) into (\ref{Hur}) gives
\be \label{S7}
A_{11}+A_{22}=-2(1-g|K|^{\alpha}) , \quad
\ee
\be \label{S8}
A_{11}A_{22}-A_{12}A_{21}=
\Bigl( b(1-g|K|^{\alpha})- (a-gc|K|^{\alpha} ) \Bigr)
\Bigl( 3b(1-g|K|^{\alpha})-(a-gc|K|^{\alpha} ) \Bigr) .
\ee
Then the conditions (\ref{Hur}) have the form
\[
1-g|K|^{\alpha}>0 ,
\]
\be \label{Stab}
1-g|K|^{\alpha} < a/b-(c/b)g |K|^{\alpha} < 3(1-g|K|^{\alpha}) ,
\ee
i.e. the plane-wave solution (\ref{B6})
is unstable if the parameters $a$, $b$, $c$ and $g$ do not satisfy (\ref{Stab}).
Condition (\ref{Stab}) defines the region of parameters 
for plane waves where the synchronization can exist.

In the numerical simulation, we use the parameters
\be
a=-1.2, \quad b=-2, \quad c=2, \quad g=1 .
\ee
Then Eq. (\ref{Stab}) gives the inequalities
\be
0< 1-|K|^{\alpha} < 0.6+|K|^{\alpha} < 3(1-|K|^{\alpha}) .
\ee

As a result, the plane-wave solution with $E=0$ is stable for
\be \label{Bound}
0.2 < |K|^{\alpha}<0.6, \quad (1<\alpha<2) .
\ee
In our numerical simulation, we use the initial conditions with
\be
|K|=2 \pi /64 \approx 0.09817
\ee 
that is $K$ is in the unstable region for $1< \alpha <2$. 
In the following the initial conditions with perturbation $E \ne 0$
will be out of the boundaries (\ref{Bound}),
and evolution of the initially unstable states will be studied.

\section{Mapping the numerical data}

Numerical results are obtained as solutions of the coupled equations  
\be \label{FGLE}
\frac{d}{dt}Z_n=(1+ia)Z_n-(1+ib)|Z_n|^2 Z_n +
\frac{1+ic}{a_{\alpha} (\Delta x)^{\alpha}} \sum^N_{m \not=n} \frac{1}{|n-m|^{\alpha+1}} (Z_n-Z_m) +E,
\ee
where $n=1,..,N$, $\Delta x=1$, and $E$ is a constant external force.
In simulations, the number of oscillators was $N=512$, and
the external force was $E=0.3$.
In the continuous limit, Eq. (\ref{FGLE}) transforms into the FGL equation
with forcing. 

For all sets of parameters, we integrate the DFGL equations with 
the initial conditions
\be
Z_n(0)=A_0 e^{iKn} ,
\ee 
where $|K|=2 \pi /T$, the space period is $T=64$, and 
$A_0$ is an initial amplitude.
Numerical solutions were stored at each $t_q = 0.005 q$, 
where $q \in\mathbb{N}$. \\

To visualize the numerical solution we consider the following values. 

(1) We plot the surface $|Z(x,t)|^2$ and the phase-space projection 
of the central oscillator ($n=0, x_n = 0$) formed by the variables:
\be
\label{Aplate}
A(t) = |Z(0,t)|^2, \quad \dot{A}(t) = \frac{dA(t)}{dt}.
\ee 

(2) In addition to (\ref{Aplate}), we plot 
$Y(t)=Im(Z(0,t))$ vs. $X(t)=Re(Z(0,t))$ for better resolution
of the central oscillator behavior. \\

(3) We calculate the discrete Fourier 
transform of the sequence $Z(0,t_q)$ defined as
\be
\label{f_time}
\hat{Z}_n(\omega_j)  = \frac{1}{Q} \sum_{q=0}^{Q-1} Z_n(t_q) \; 
\exp(- i \; \omega_j \; t_q),
\ee
\be
\label{f2_time}
Z_n (t_q) = \sum_{j=0}^{Q-1} \hat{Z}_n (\omega_j) \; \exp(i \; \omega_j \; t_n), 
\ee
where $\omega_j=2 \pi j/Q$, ($j=0, ..., Q-1$). 
The power spectrum $S$ of the sequence $Z_n(t_q)$ for $q=0, ..., Q-1$ is given by 
\be
\label{Splate}
S_j \equiv S(\omega_j) = |\hat{Z}_n (\omega_j)|^2.
\ee
Our main goal is to compare solutions of the DFGL equation for 
different values of $\alpha \in (1,2)$ and consider emergence of 
chaotic dynamics of the chain of oscillators as a function of $\alpha$. 
The larger is $\alpha$, the weaker is long-range interaction. 

In the simulations, we consider the following plots. \\
(a) Color plots present surfaces $|Z_n(t)|^2$ vs $t$ and $n$; \\
(b) Plane ($A, dA/dt$) displays a projection of the trajectory of the 
central oscillator (see definition in (\ref{Aplate})); \\
(c) Plane ($Re\ Z, Im\ Z$) displays a projection of 
the complex amplitude $Z=Z(0,t)$ 
of the central oscillator as a function of time; \\
(d) Plots $(\log_{10} \omega,\log_{10} S)$ describe the spectrum of 
time oscillations of $Z(0,t)$ (see definition in Eq.\ (\ref{Splate})).


\section{Some numerical results for CGL equation}

In this section, we provide some numerical results derived in 
\cite{Ch,Heck,CPR,Pik2,TZ3} for the complex Ginzburg-Landau (CGL) equation
in order to compare them, obtained for $\alpha=2$, 
with our results for $\alpha <2$.
For many other results and details see \cite{AK,Pik1,Z1}. 

(a) In paper \cite{Ch}, the CGL equation has been considered for the parameters
\be
\alpha=2, \quad a=0 \quad b=1.333, \quad c=-1 .
\ee 
and the phase turbulence has been observed.

(b) It was shown in \cite{Heck} that for the parameters
\be
\alpha=2, \quad a=0, \quad b=-1.4, \quad c=0.6 
\ee
solution of the CGL equation has chaotic states 
and it is spatial-temporal intermittent.

(c) It was found in \cite{Heck} that for the parameters
\be
\alpha=2, \quad a=0, \quad b=1.2, \quad c=-0.6 
\ee
solution of the CGL equation has zigzagging holes 
near the transition to the plane waves.

(d) In \cite{CPR}, the CGL equation was considered 
for the parameters
\be \label{d-d}
\alpha=2, \quad a=-1.2, \quad b=-2,\quad c=2 .
\ee
For $E=0.35$ the solution displays pitchforks without defects 
while for $E=0.23$ the rare defects were observed.  

(e) In \cite{TZ3}, the FGL equation was considered for the parameters
\be 
1<\alpha<2, \quad a=1, \quad b=0, \quad c=70 .
\ee
It was shown that the solution had one stable fixed point for 
$\alpha_0< \alpha<2$, ($\alpha_0=1.51$)
that corresponds to synchronization of oscillators.
Decreasing of $\alpha$ below $\alpha_0$ leads to a limit cycle 
via the Hopf bifurcation.


\section{Transition from synchronization to turbulence}


Regular propagation in time of the initial state of the chain
of oscillators will be called synchronization. 
To consider $\alpha$-dependence of transition from 
synchronization to turbulence near $\alpha=2$, we use the parameters
\be \label{43}
a=-1.2, \quad b=-2, \quad c=2 ,\quad A_0=0.2 ,
\ee
similar to (\ref{d-d}) that were used in  \cite{CPR}
for the CGL equation (\ref{A2}), but for the chain of oscillators with
$\alpha \ne 2$.
The simulation was performed for the chain of 512 oscillators
with parameters equivalent to (\ref{43}). \\


\begin{figure}
\centering
\rotatebox{0}{\includegraphics[width=12 cm,height=18 cm]{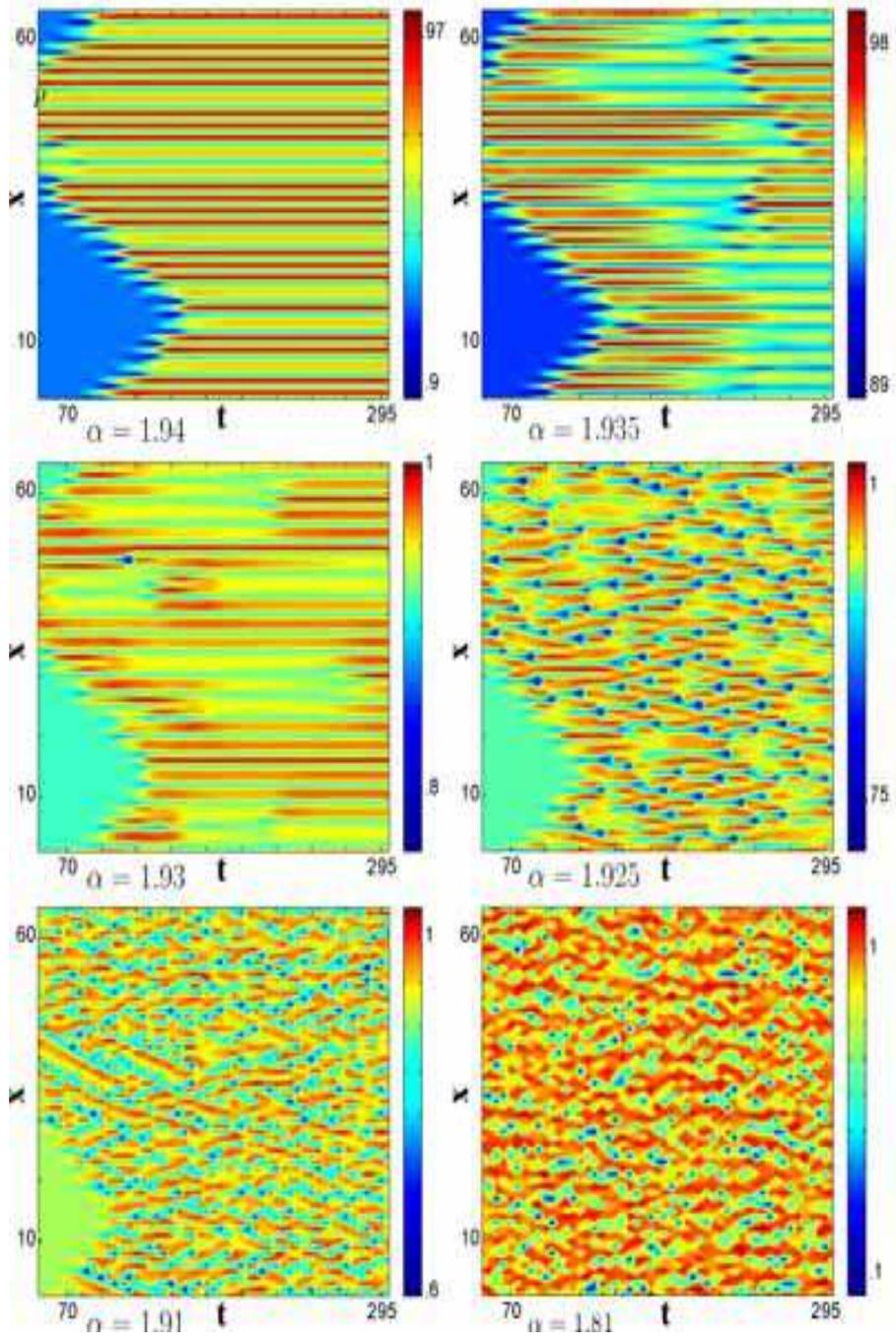}}
\caption{\label{Fig1}  
Alpha-dependence of transition from synchronization 
to turbulence near $\alpha=2$. 
Surfaces of $|Z_n(t)|^2$ vs $t$ and $x=n$ for
$\alpha=1.940$, $\alpha=1.935$, $\alpha=1.930$, 
$\alpha=1.925$, $\alpha=1.910$,  $\alpha=1.810$.
Simulations are realized for FGL equation with parameters
$a=-1.2$, $b=-2$, $c=2$, $A_0=0.2$. }
\end{figure}

In Fig.1., 
the surfaces of $|Z_n(t)|^2$ vs $t$ and $x=n$ for
$\alpha=1.940$, $\alpha=1.935$, $\alpha=1.930$, 
$\alpha=1.925$, $\alpha=1.910$,  $\alpha=1.810$ are shown.
For $\alpha=1.940$, we have a regular space structure of the plane-wave type.
For $\alpha=1.935$, there is a space modulation, which deforms the regular space structure. 
For $\alpha=1.930$, we can see the appearance of defects and pitchforks.
For $\alpha=1.925$, the structure in the space-time demonstrates sharp and drastic changes. 
There exist many defects and pitchforks.
For $\alpha=1.910$, the number of defects and pitchforks is increased,
and some pitchforks are joined.
For $\alpha=1.810$, we can see  chaos and turbulence, and 
synchronization is completely lost.
We see that the amplitude turbulence 
is characterized by persistent creation and annihilation of pitchforks.
The decreasing of order of fractional derivative 
means increasing a role of LRI. It is worthwhile to comment that
the loss of synchronization and the emergence of amplitude turbulence 
is fairly sharp with fairly small change of $\alpha$. \\


\begin{figure}
\centering
\rotatebox{0}{\includegraphics[width=15 cm,height=15 cm]{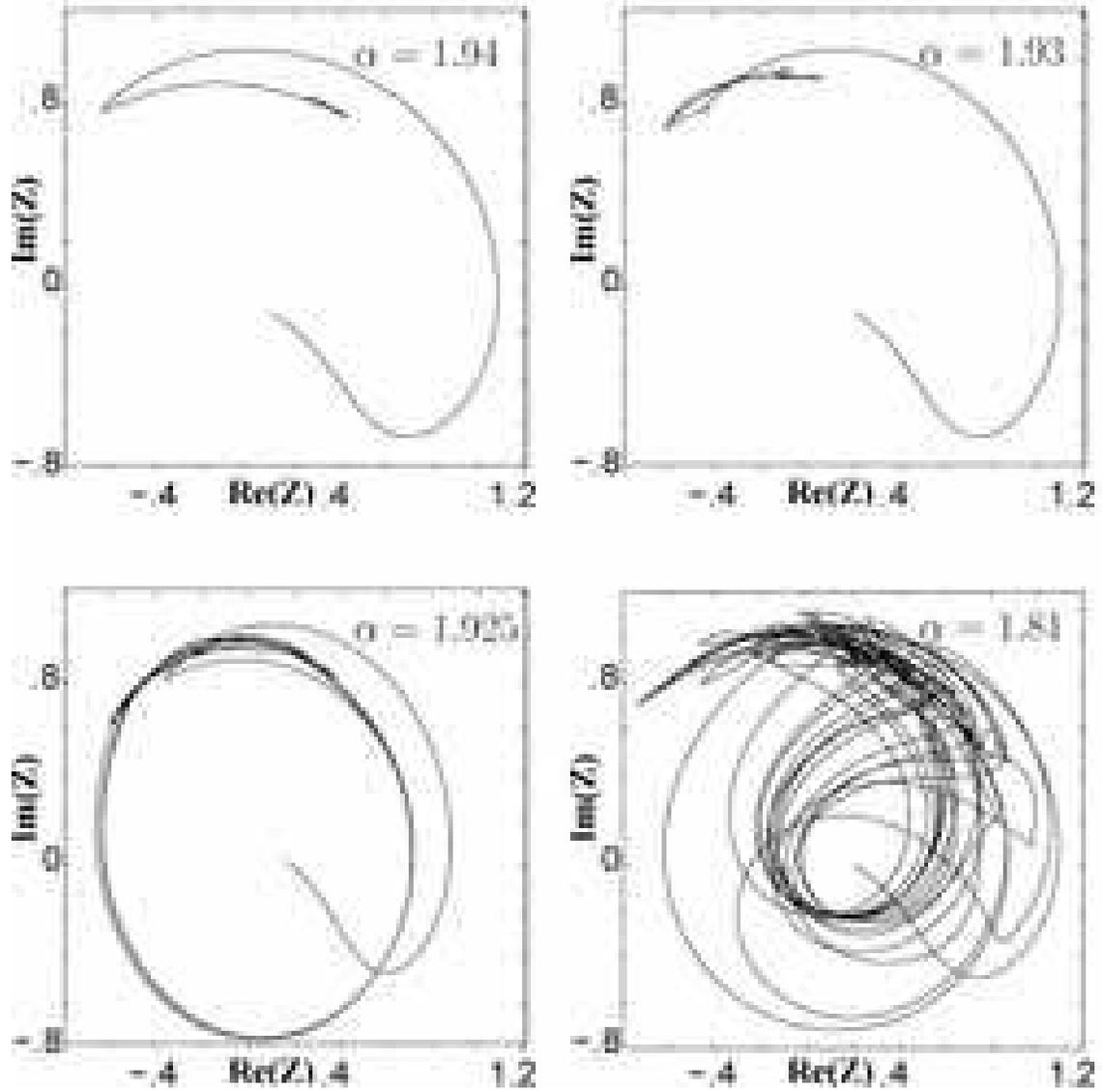}}
\caption{\label{Fig2} Alpha-dependence of transition from synchronization 
to turbulence near $\alpha=2$. 
Plane ($Re\ Z, Im\ Z$) shows projection of the complex amplitude $Z=Z(0,t)$ 
of the central oscillator as a function of time  for
$\alpha=1.940$,  $\alpha=1.930$, $\alpha=1.925$, $\alpha=1.810$.}
\end{figure}

\begin{figure}
\centering
\rotatebox{0}{\includegraphics[width=15 cm,height=15 cm]{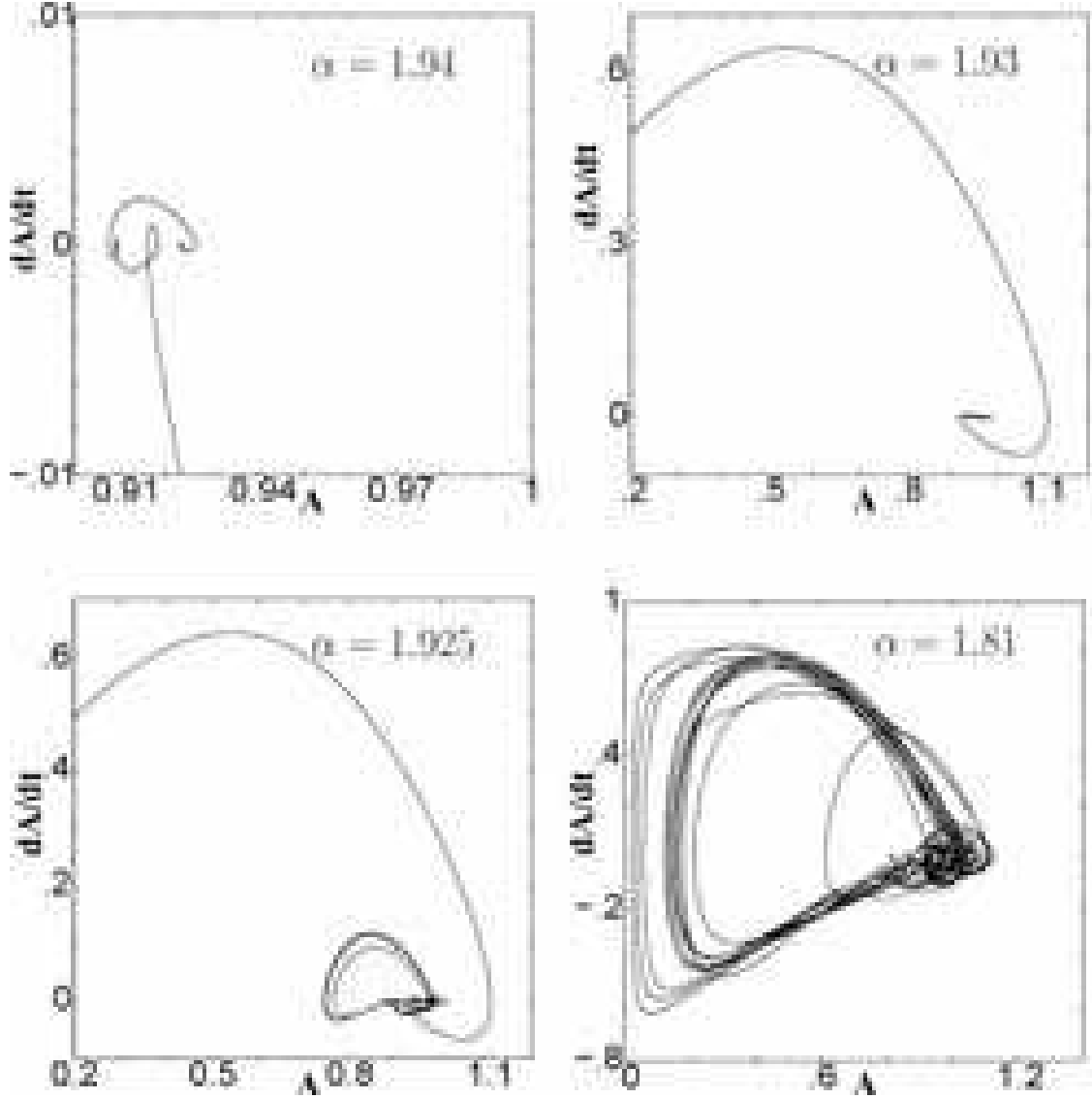}}
\caption{\label{Fig3}  
Alpha-dependence of transition from synchronization 
to turbulence near $\alpha=2$. 
Plane ($A, dA/dt$) shows projection of the trajectory of the 
central oscillator in phase space, where $A(t) = |Z(0,t)|^2$
for $\alpha=1.940$,  $\alpha=1.930$, $\alpha=1.925$, $\alpha=1.810$.}
\end{figure}

In Fig.2.,
the plane ($Re\ Z, Im\ Z$) shows projection of the complex amplitude $Z=Z(0,t)$ 
of the central oscillator as a function of time  for
$\alpha=1.940$,  $\alpha=1.930$, $\alpha=1.925$, $\alpha=1.810$.
For $\alpha=1.94$ and $\alpha=1.93$, there is a stable node, which 
means the existence of synchronization. 
Conformation of that can also be found in Fig.3, where 

the plane ($A, dA/dt$) shows projection of the trajectory of the 
central oscillator and $A(t) = |Z(0,t)|^2$.
For $\alpha=1.94$ and $\alpha=1.93$, the attracting point that  
corresponds to synchronization of the chain of oscillators is clearly seen. 
Characteristics of the turbulent motion, shown in Fig.1. for $\alpha=1.925$,
 $\alpha=1.91$,  $\alpha=1.81$, are different from the cases of larger $\alpha$.
This difference can be better recognized from the Figs.2-4.
Particularly, in  Fig.2. for $\alpha-1.925$  and  $\alpha=1.81$ behavior of  
$Re Z=ReZ(0,t)$ and $Im Z=ImZ(0,t)$ displays a disordered process 
that on the plane $(A,dA/dt)$ in Fig.3.
reveals a structure similar to what is usually observed for stochastic attractors.
Nevertheless, more specific statement needs more detailed analysis since
our system is open and has many degrees of freedom.


\begin{figure}
\centering
\rotatebox{0}{\includegraphics[width=15 cm,height=15 cm]{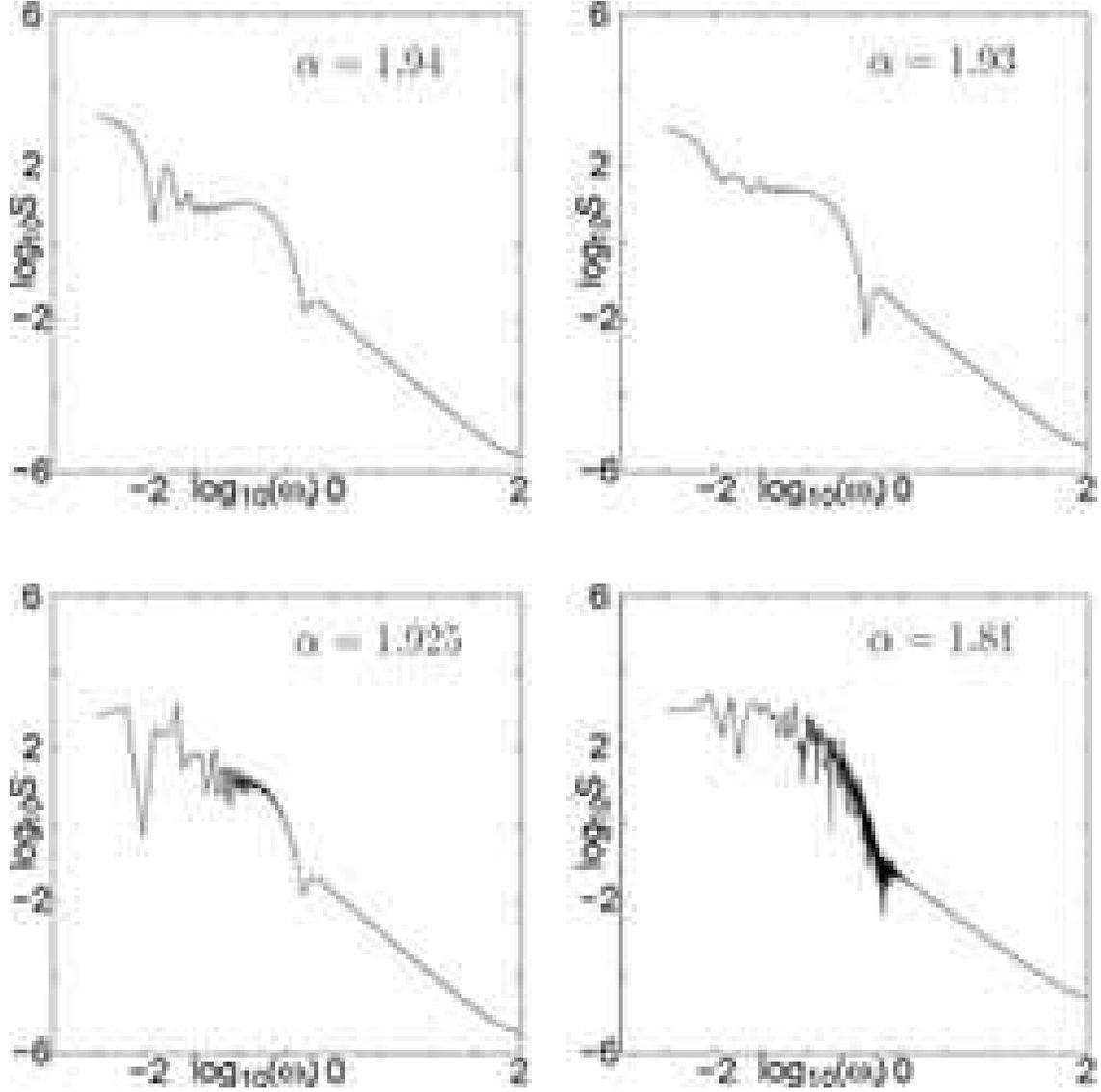}}
\caption{\label{Fig4} Alpha-dependence of transition from synchronization 
to turbulence near $\alpha=2$. 
Spectrum of time oscillations of $Z(0,t)$ (see definition in Eq.\ (\ref{Splate}))
for $\alpha=1.940$,  $\alpha=1.930$, $\alpha=1.925$, $\alpha=1.810$. }
\end{figure}

In Fig.4., we show 
the spectrum of time oscillations of $Z(0,t)$ (see definition in Eq.\ (\ref{Splate}))
for $\alpha=1.940$,  $\alpha=1.930$, $\alpha=1.925$, $\alpha=1.810$.
For $\alpha=1.94$ and $\alpha=1.93$, the spectrum has small numbers of harmonics,  
and it corresponds to the regime of synchronization in the oscillator medium. 
For $\alpha=1.925$, the spectrum is filled by additional harmonics.
For $\alpha=1.81$, the spectrum has a dense set of frequencies 
for the region $0.1< \omega<1$ that 
reflects the chaos behavior and turbulence of the oscillator medium. 
For the region $1<\omega<100$, we can see the integer power-law for the spectrum
$S(\omega) \sim \omega^{-2}$ that corresponds to the main frequency dependence
of equation (\ref{A8}).


\section{Traveling waves and broadening of the limit cycle}

Large number of oscillators and periodic boundary conditions permit 
to consider localized traveling waves (see for example 
for discrete systems in \cite{Z10,Z11}).
Such waves were observed in our system (\ref{FGLE})
near the values of $\alpha \stackrel{<}{\sim} 2$ and disappear near $\alpha \approx 1$.
In the latter case, it will be also demonstrated space-temporary 
localized topological defects.

The simulation was performed for the parameters 
\be \label{par2}
a=1, \quad b=0, \quad c=2 , \quad A_0=0.2.
\ee
For the value $b=0$, the nonlinear term is real.

The waves propagation can be characterized by the group velocity 
$v_{\alpha,g}={\partial \omega_{\alpha}(K)}/{\partial K}$.
From Eq. (\ref{B7}), we obtain
\be
v_{\alpha,g} =\alpha (c-b) g |K|^{\alpha-1}. 
\ee
The phase velocity is 
\be
v_{\alpha,ph}= \omega_{\alpha}(K) / K=\frac{b-a}{K} +(c-b)g |K|^{\alpha-1} .
\ee
For the parameters used in the simulations
we have $|v_{\alpha,g}| > | v_{\alpha,g}|$.
In our simulation $K=2\pi /64$.
For our simulation $K=2\pi /64$, and 
the decreasing of the order $\alpha$ leads to 
the increasing of the group and phase velocities. \\


\begin{figure}
\centering
\rotatebox{0}{\includegraphics[width=12 cm,height=17 cm]{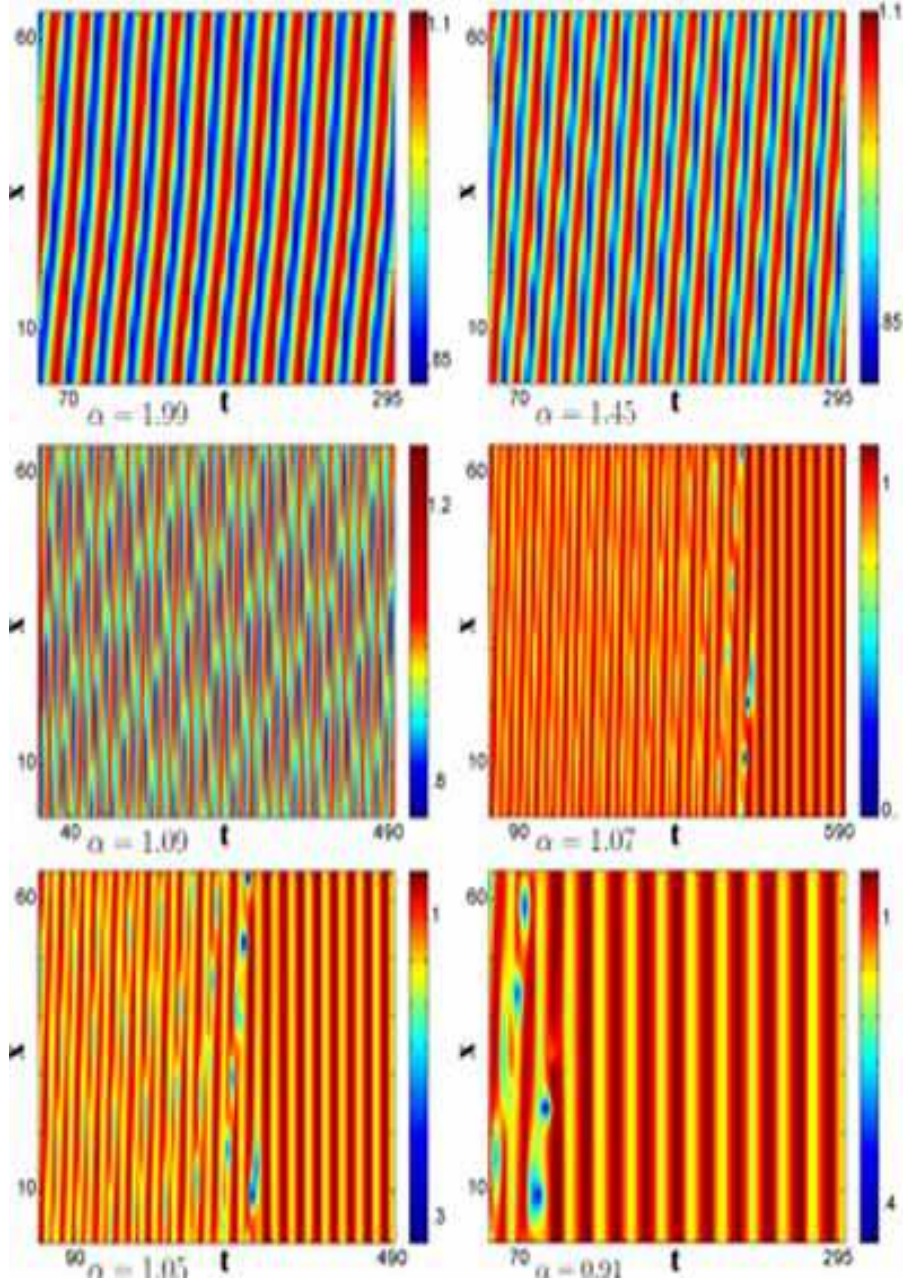}}
\caption{\label{Fig5} 
Broadening of limit cycle. Wave front deinclination.
Surfaces of $|Z_n(t)|^2$ vs $t$ and $x=n$ for
$\alpha=1.99$, $\alpha=1.45$, $\alpha=1.09$, 
$\alpha=1.07$, $\alpha=1.05$,  $\alpha=0.91$.
Simulations are realized for FGL equation with parameters
$a=1$, $b=0$, $c=2$, $A_0=0.2$. }
\end{figure}

\begin{figure}
\centering
\rotatebox{0}{\includegraphics[width=12 cm,height=17 cm]{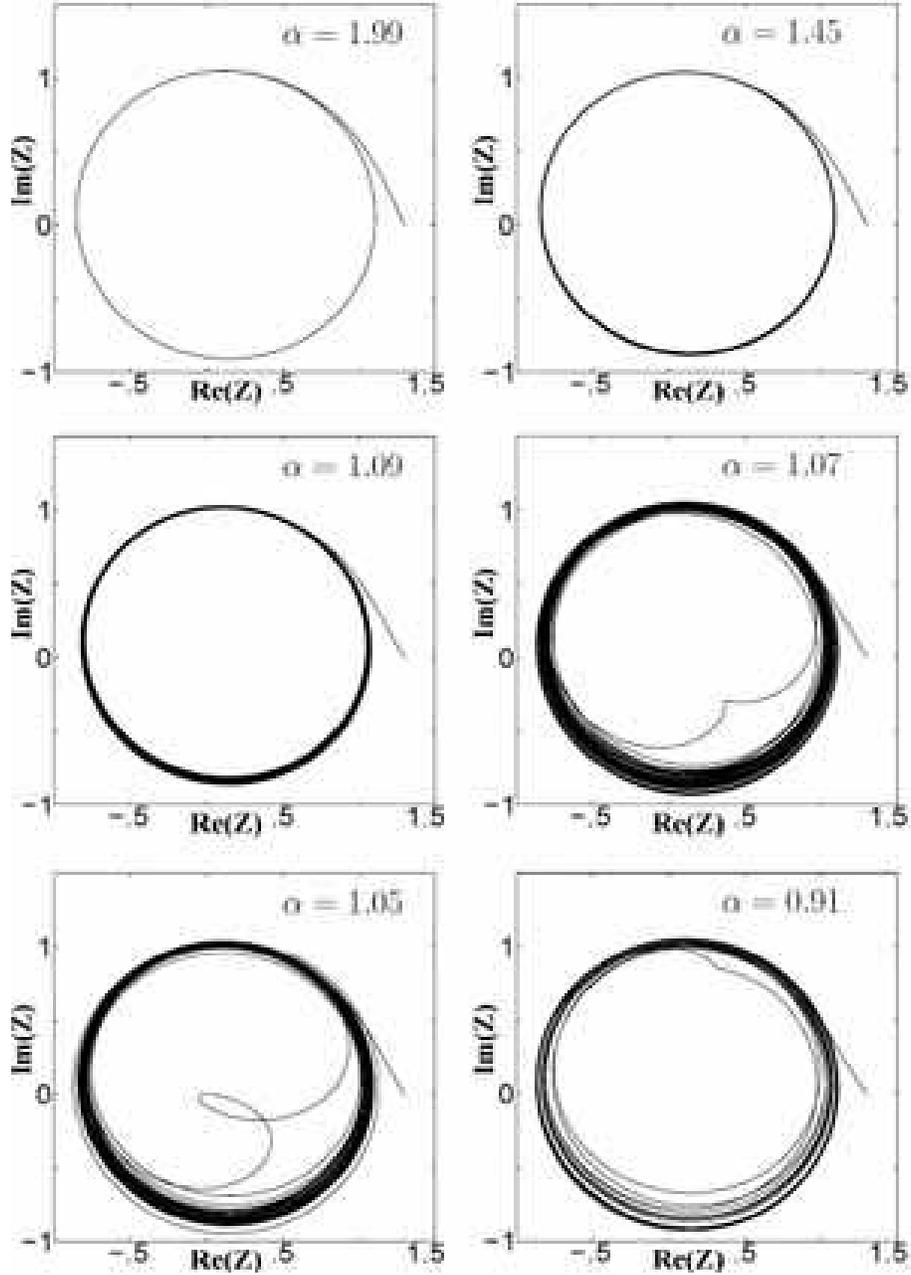}}
\caption{\label{Fig6} Broadening of limit cycle.
Plane ($Re\ Z, Im\ Z$) shows projection of the complex amplitude $Z=Z(0,t)$ 
of the central oscillator as a function of time  for
$\alpha=1.99$, $\alpha=1.45$, $\alpha=1.09$, 
$\alpha=1.07$, $\alpha=1.05$,  $\alpha=0.91$.}
\end{figure}

\begin{figure}
\centering
\rotatebox{0}{\includegraphics[width=10 cm,height=14 cm]{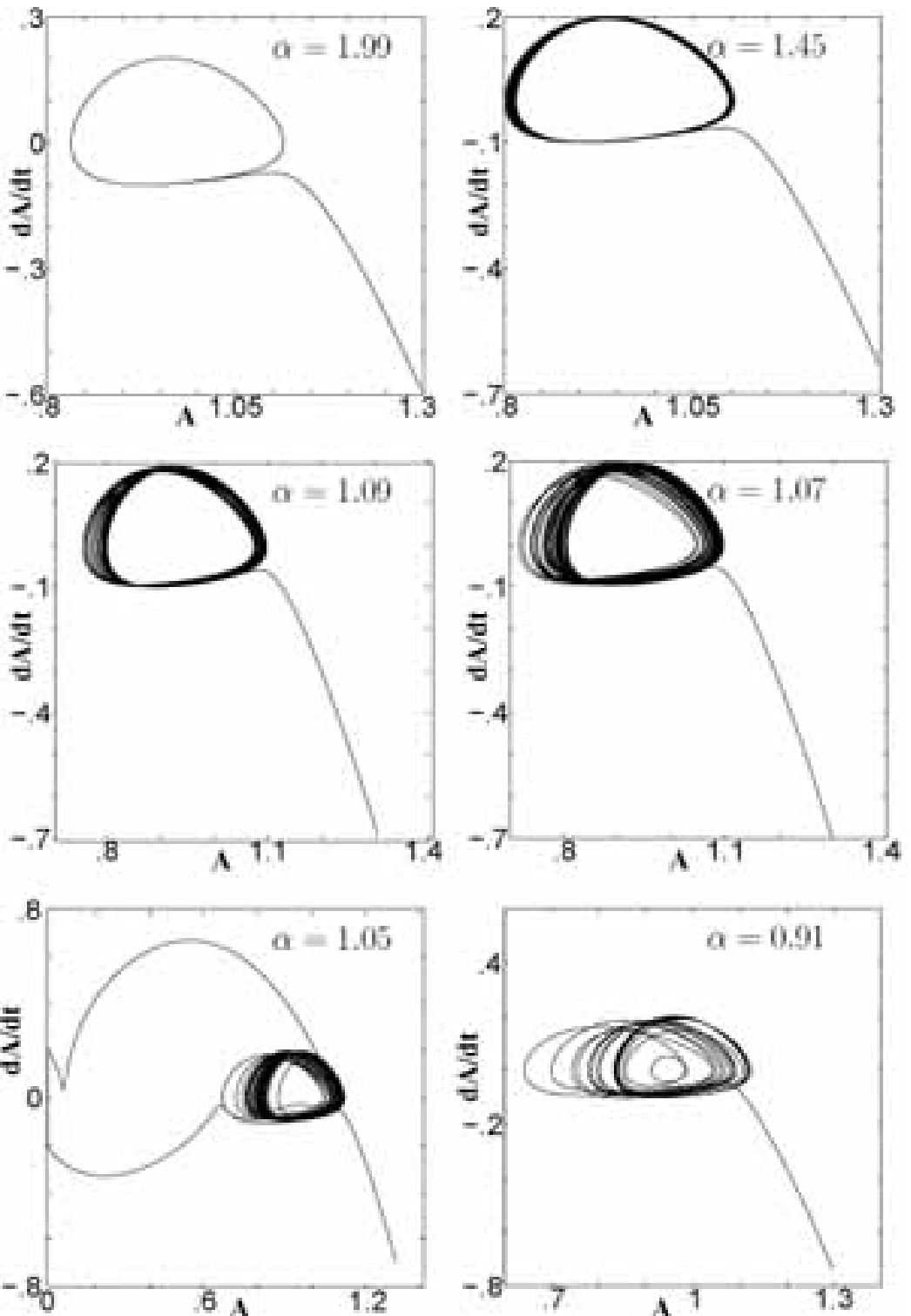}}
\caption{\label{Fig7}  Broadening of limit cycle. Appearence of topological defect. 
Plane ($A, dA/dt$) shows projection of the trajectory of the 
central oscillator in phase space, where $A(t) = |Z(0,t)|^2$
for $\alpha=1.99$, $\alpha=1.45$, $\alpha=1.09$, 
$\alpha=1.07$, $\alpha=1.05$,  $\alpha=0.91$.}
\end{figure}

The corresponding simulation is shown in Figs.5-8 for 
the surfaces $|Z_n(t)|^2$ (Fig.5), $ReZ$ and $ImZ$ plane (Fig.6),
$(dA/dt,A)$ - plane (Fig.7) and spectrum $S(\omega)$ in Fig.8.
Each of the the plate has the some additional information about solutions.
There is a strong difference between values $\alpha=1.99$,
$\alpha=1.45$, and $\alpha=1.07$, $\alpha=1.05$,  $\alpha=0.91$ 
while the case of $\alpha=1.09$ is not clear since 
the finite time of simulation ($t<1000$).
The first case ($\alpha=1.99$, $\alpha=1.45$) shows traveling waves 
along $x$ with regular (periodic or quasi-periodic) patterns.
In Figs~5,6, we observe the approach of the solution 
to a limit cycle with  a few harmonics in the spectrum (Fig.8).
At the same time the smaller is $\alpha$,
the smaller scales (larger values of $K$) enter the solution.

For $\alpha$ close to one ($\alpha=1.07$, $\alpha=1.05$,  $\alpha=0.91$)
a fairly irregular pattern of traveling wave declines at some time, and
synchronized oscillations take place.
In the phase diagrams in Figs~6,7 a broadened limit cycle type picture
corresponds to collective oscillations of the chain with fairly rich
spectrum presented in Fig.8.
The closer is $\alpha$ to one, the shorter is time of break of traveling
waves, and the synchronized breather type solution appears.
Since the growth of wave numbers $K$ of the solution, 
its amplitude can reach zero giving rise the topological defects \cite{AK}.
Their appearance is clearly seen from Figs~6,7 for $\alpha =1.05$ (see
also Fig.5), when the amplitude  $A(t) = |Z(0,t)|^2$ reaches zero value.
It is seen that for $\alpha=1.07$ and  $\alpha=0.91$
the dynamics is close to the appearance of the topological defect. 
Zero of the complex field $Z$ result in singularity of
the phase $\theta= arg Z$. In two dimensions, points of singularities
correspond to quantized vortices with topological charge
\be \label{47}
  n =\frac{1}{2\pi} \oint_L  \nabla \theta dl,
\ee
where $L$ is a contour encircling a zero of $Z$. 

For $1<\alpha<2$, we have the limit cycles around 
the point $A=1$, $dA/dt=0$. 
It is easy to see the broadening of 
these limit cycles, when $\alpha$  decreases. \\


\begin{figure}
\centering
\rotatebox{0}{\includegraphics[width=12 cm,height=18 cm]{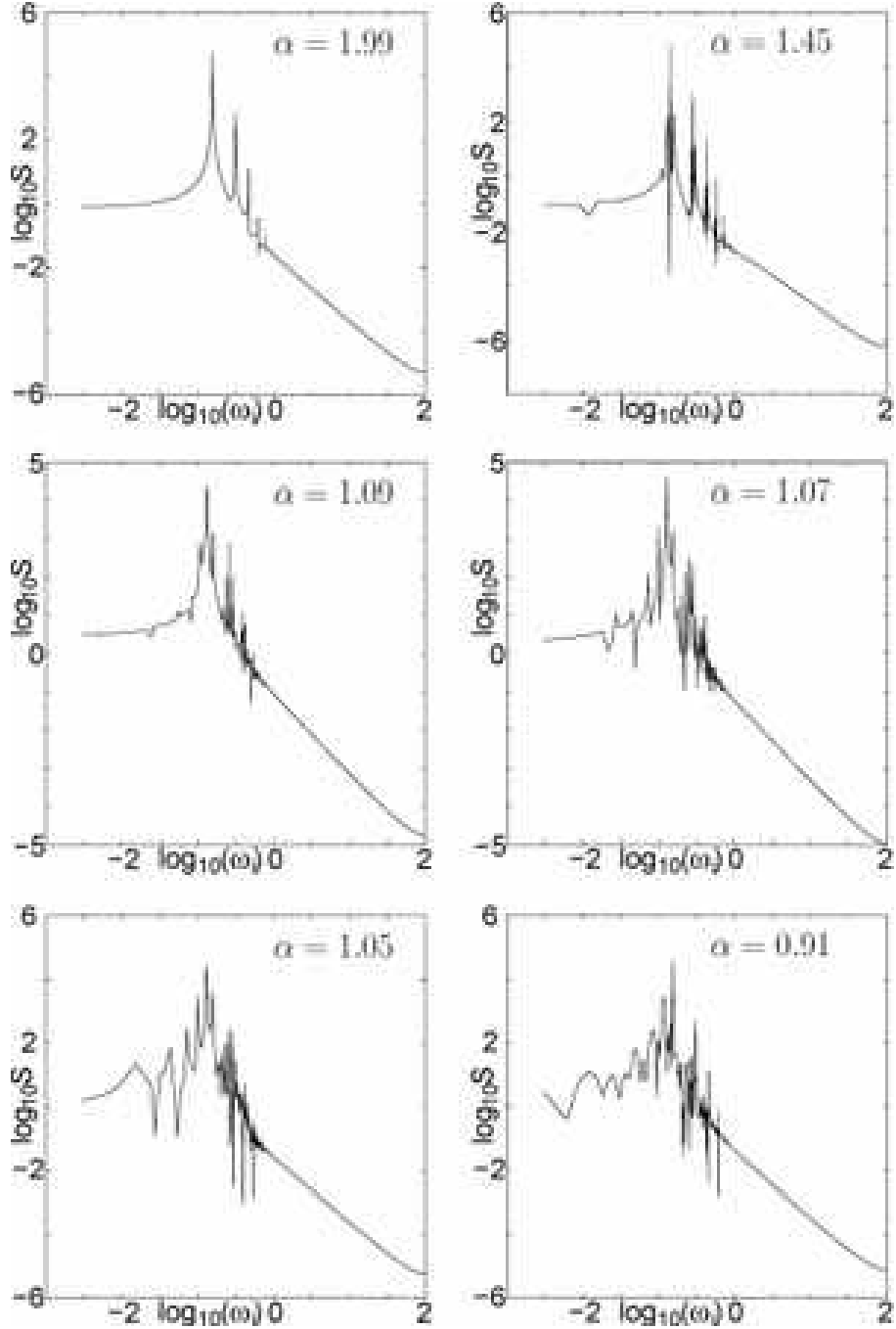}}
\caption{\label{Fig8} Broadening of limit cycle. 
Spectrum of time oscillations of $Z(0,t)$ (see definition in Eq.\ (\ref{Splate}))
for $\alpha=1.99$, $\alpha=1.45$, $\alpha=1.09$, 
$\alpha=1.07$, $\alpha=1.05$,  $\alpha=0.91$.}
\end{figure}

Finally for this section, let us  comment on
the spectrum of time oscillations of $Z(0,t)$ 
for different $\alpha$, presented in Fig.8.
The broadening of limit cycle leads to widening of the spectrum.
The spectrum is filled out by different harmonics
that are localized in the region $0.1<\omega<1$. 
For $\omega > 1 $, we have the dependence
$S(\omega) \sim \omega^{-2}$ that follows 
directly from equation (\ref{A2}).


\section{Amplitude dependence of the transition to turbulence}

In this section, we would like to show that the turbulent regime
developing depends on the initial amplitude. 
For simulation, we use some parameters
as in (\ref{d-d}) \cite{CPR} except of $\alpha=1.45$.


\begin{figure}
\centering
\rotatebox{0}{\includegraphics[width=14 cm,height=7 cm]{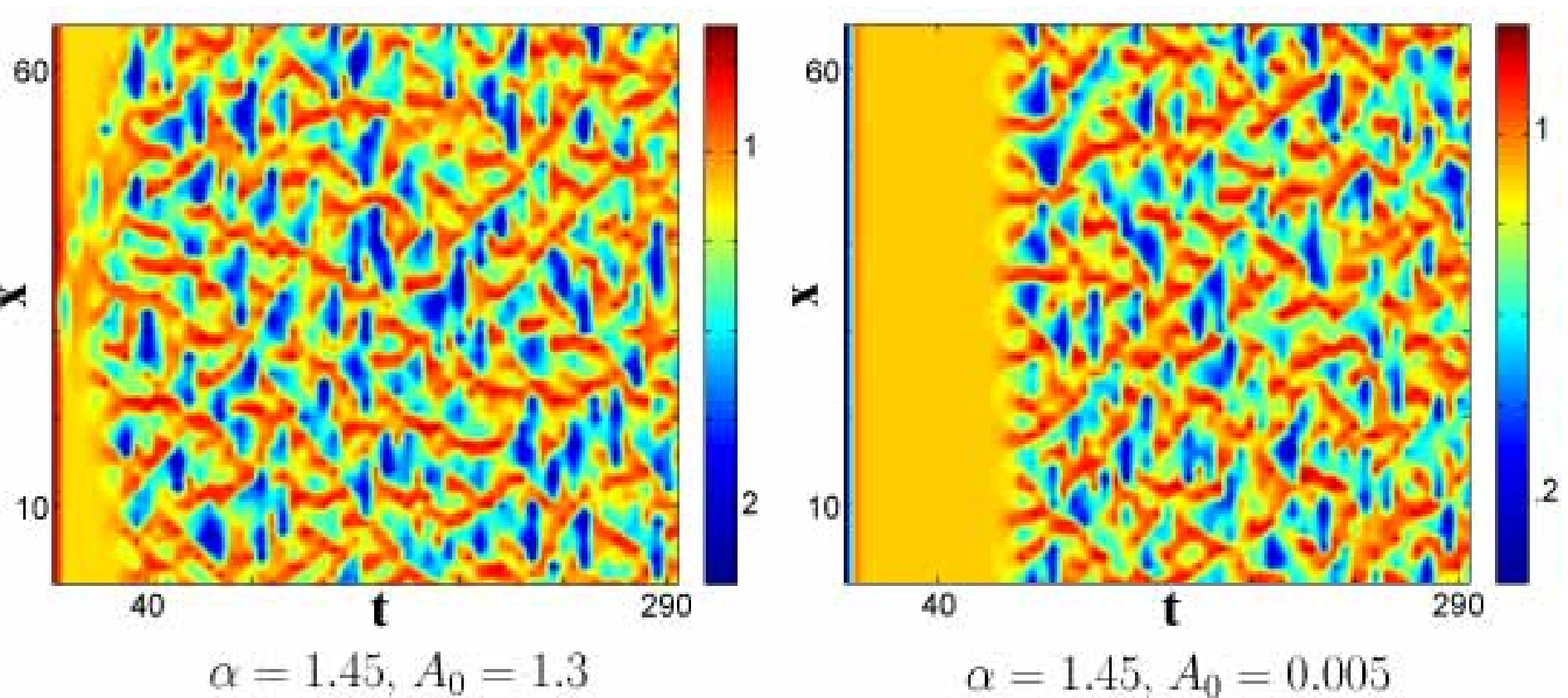}}
\caption{\label{Fig9}  Amplitude dependence of transition to turbulence.
The decreasing of the initial amplitude from $1.3$ to $0.005$ gives that
the space-time structure of turbulence almost is not changed.
There is only increasing of time of begining of turbulence.}
\rotatebox{0}{\includegraphics[width=14 cm,height=7 cm]{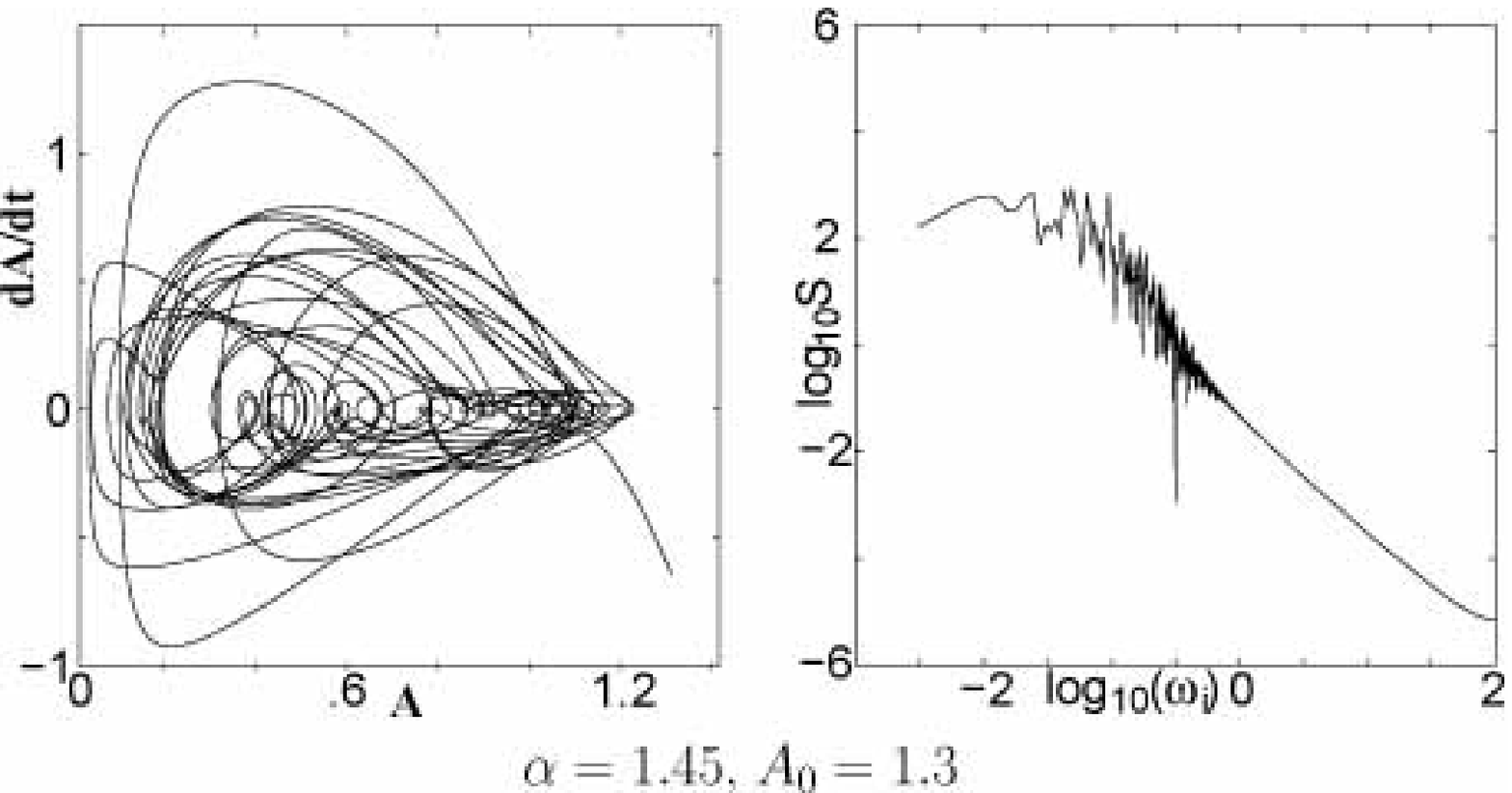}}
\caption{\label{Fig10} Amplitude dependence of transition to turbulence. 
Plane ($A, dA/dt$) shows projection of the trajectory of the 
central oscillator in phase space, where $A(t) = |Z(0,t)|^2$, and
the spectrum of time oscillations of $Z(0,t)$ for $\alpha=1.45$ and $A=1.3$.}
\end{figure}

It is seen from Fig.9, that 
the decreasing of the initial amplitude $A_0$ from $1.3$ to $0.005$ 
doesn't change, at least visually, 
the space-time structure of turbulence,  
but the change increases the time of developing an instability and 
beginning of turbulence.
The turbulence possesses a robustness with respect 
to change of the amplitude of initial excitation and initial conditions. \\


In Fig.10., 
the plane ($A, dA/dt$) shows projection of a trajectory of the 
central oscillator. There are random rotations  
on the plane and a broadening of the spectrum in a finite
region of frequencies. 
The Fig.~10 shows many loops of the trajectory for which $dA/dt=0$ and $A$
is close to zero. That means that the turbulent regime includes many
different points that are close to be the topological defects.


\section{Conclusion}

The main goal of this paper was to study influence of 
long-range interaction (LRI) on the developing of 
chaotic or turbulent motion in one-dimensional chain
of large number of nonlinear oscillators.
The LRI is characterized by the power of interaction $\alpha$.
In the continuous limit the corresponding equation is the
nonstationary generalized Ginzburg-Landau (FGL) equation 
with complex coefficients and fractional derivatives of order $\alpha$
along the coordinate variables.
Our preliminary research show different interesting regimes 
of behavior of the chain depending on the value of $\alpha$.
We have observed a synchronized motion of the chain 
with different complexifications such as defects, chaos,
space-time turbulence, traveling waves.
A possibility to use a continuous analog of 
the chain such as FGL equation, 
simplifies some estimates although it is well-known 
that discrete model is "more chaotic" than the continuous one.

\section*{Acknowledgments}

This work was supported by the Office of Naval Research, 
Grant No. N00014-02-1-0056, and the NSF Grant No. DMS-0417800.

\end{document}